\documentclass{JAC2000}

\usepackage{graphicx}

\begin{document}

\title{BEAM SWITCHING AND BEAM FEEDBACK SYSTEMS AT KEKB LINAC}

\author{
K. Furukawa\thanks{e-mail: \texttt{<kazuro.furukawa@kek.jp>} }, 
A. Enomoto, N. Kamikubota, T. Kamitani, T. Matsumoto,\\
Y. Ogawa, S. Ohsawa, K. Oide and T. Suwada \\
High Energy Accelerator Research Organization (KEK) \\
Oho 1-1, Tsukuba, Ibaraki, 305-0801, Japan}

\maketitle

\begin{abstract} 

The KEK 8-GeV electron / 3.5-GeV positron linac has been operated with
very different beam specifications for downstream rings, KEKB, PF and
PF-AR.  For the reliable operation among these beam modes, intelligent
beam switching and beam feedback systems have been developed and used
since its commissioning.  

A software panel is used to choose one of four beam modes and a
switching sequence is executed in about two minutes. Most items in a
sequence are simple operations followed by failure recoveries.  The
magnet standardization part consumes most of the time. The sequence
can be easily re-arranged by accelerator operators. Linac beam modes
are switched about fifty times a day using this software.

In order to stabilize the linac beam energy and orbits, as well as
some accelerator equipment, about thirty software beam feedback loops
have been installed.  They have been routinely utilized in all beam
modes, and have improved its beam quality.  Since its software
interfaces are standardized, it is easy to add new feedback loops
simply defining monitors and actuators.  


\end{abstract}

\section{Introduction}

The KEK electron/positron linac had been upgraded for the KEK (KEK
B-factory) asymmetric electron-positron collider since
1994. Commissioning of the first part of the linac started at the
end of 1997 and has already achieved the designed beam parameters after
its completion in 1998\cite{kekb-linac-epac2000}.  It has been
providing beams for the B-physics experiment (Belle) of the
CP-violation study at KEKB since 1999.

The performance of the experiment depends on the integrated luminosity
at KEKB, which is largely dependent on the stability and intensity of
the linac beams.  Since the linac must provide four beam modes which
are very different (KEKB $e^+$, KEKB $e^-$, PF-Ring, PF-AR), it had
been realized that it was important to achieve reproducibility and
stability of each of those four beam modes\cite{kekb-linac-linac2000}.

\section{Linac Controls}

The linac control system was also upgraded\cite{lin-cont-ical99} to
support the upgraded high-intensity linac based on the system
rejuvenation in 1993\cite{lin-cont-ical91}. It consisted of layered
components that communicate with each other, where hardware and lower-layer
information were hidden from the upper layer and only useful features
are exposed to the upper layers. New components were added to
accommodate new accelerator equipment and features for the KEKB
injection. Especially, software for beam position-monitors was
developed and the database for equipment and beam lines was much enriched.

In the commissioning, many pieces of application software were
developed as clients to the control system. Many of them were designed
with a user interface on X-Window employing SAD-Tk or tcl/tk scripting
languages for rapid development and simple manipulation. They use 
common library routines to facilitate maintenance as well as
development. The number of application programs exceed 100, including
those for beam studies.

\section{Commissioning}

In commissioning of the upgraded linac, the quality of beams had
gradually improved as the beam study advanced, and the design values were
achieved for the short term. It was, however, realized that much effort
was required to reproduce the quality and to maintain it for a longer
period.

One of the main reasons was switching between four quite different beam
modes.  The other was short-term instabilities and long-term drifts of
the equipment parameters\cite{kekb-linac-linac2000}. 

In order to cure these, the software for beam-mode switching and feedback
loops has been refined, while they had been developed since the
beginning of the commissioning.


\section{Software}

The software has been developed with the tcl/tk scripting language
under the same environment as other application
software\cite{lin-cont-ical99}.

\subsection{Beam Mode Switch}

In linac beam mode switching, as described above, 
it is important to select the operation conditions and parameters of 
the accelerator equipment reliably, and to achieve reproducible 
beam qualities.  In order to meet this purpose, the software was 
developed to be easily re-arranged, 
and it currently has the following switching items:

\begin{figure}[b]
\centering
\includegraphics*[width=60mm]{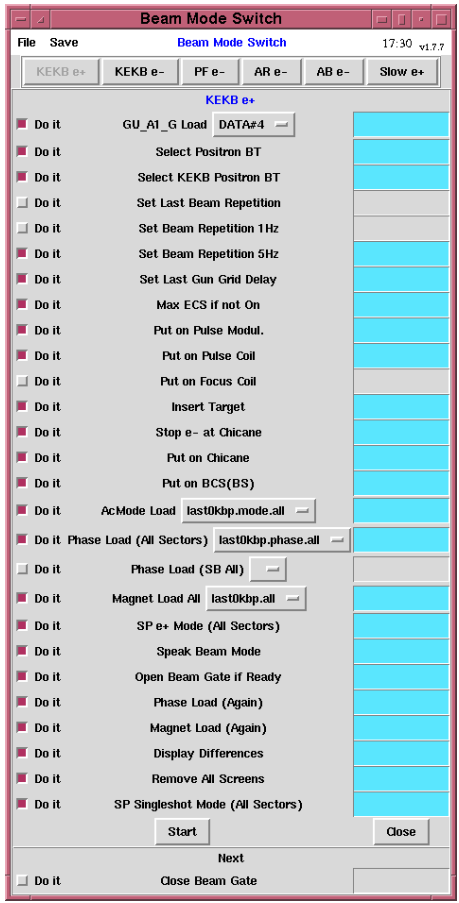}
\caption{An example of the linac beam mode switch panel. 
In this example, KEKB $e^+$ injection was selected.  
The check-buttons on the left are used to select items to 
go through.  Pull-down-menus are used to choose parameter files. 
The boxes on the right show the execution status. }
\label{fig-swi}
\end{figure}

\begin{Itemize}
\item  Suspension of beam feedback loops and other sub-systems. 
\item  De-gauss of a bending magnet (only for PF injection)
\item  Simple standardization of magnets.
\item  Selection of a gun, magnets, and rf systems.
\item  Parameters for magnets (mostly magnetic fields).
\item  Parameters for rf systems (mostly phases).
\item  Parameters for timing systems.
\item  Parameters for guns.
\item  Operation on positron targets and chicane.
\item  Operation mode of beam instrumentations and their dynamic ranges.
\item  Initial beam profile monitor selection. 
\item  Initial beam repetition rate. 
\item  Selection of beam transport lines.
\item  Information to downstream ring control systems.
\item  Review of equipment parameters.
\item  Display and record of equipment status and parameter differences. 
\item  Resumption of corresponding beam feedback loops. 
\item  Information to operators via a speech processor. 
\end{Itemize}

Items related to the radiation safety are not included, and are
handled by a separate safety-interlock system. 

Fig.~\ref{fig-swi} shows an example of the software panel. Each item
on the panel can be enabled or disabled by any operator, and its
status can be saved or restored.  New items can be introduced by
adding entries to the database. If some problems occur, which cannot
be removed through the control system, that event is reported to the
operator, who may retry it after the problem has been removed.

The items listed as `Parameters' are normally taken from the equipment 
parameters when the last time the same beam mode was used, while 
other parameter sets can be chosen from the menu if a operator 
needs one. 

For the initialization of the magnets following issues are repeatedly 
tested: reproducibility of magnetic fields, 
tolerance of the power supplies to the steep current changes and
failure recoveries in control and application software.
Since this part consumes most of the time in the switch, it is 
still being improved. 

\subsection{Beam Feedback Loops}

Software feedback loops installed in the linac are categorized into 
three groups: stabilization for equipment parameters, the beam energy
and the beam orbit.  Their basic software structure is the same and is 
built of following parts:

\begin{figure}[b]
\centering
\includegraphics*[width=68mm]{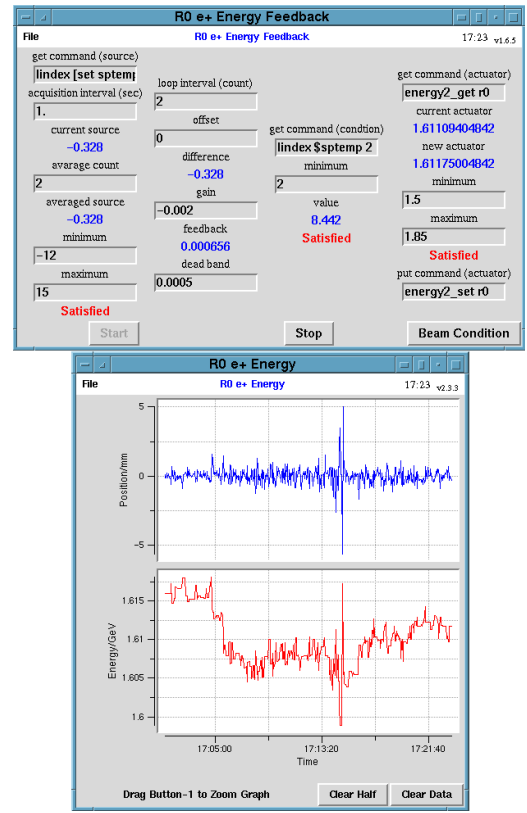}
\caption{Energy feedback panel at the R sector as an example. 
The parameters and processing specifications can be modified anytime.}
\label{fig-efb}
\end{figure}

\begin{Itemize}
\item  Check the conditions of beam modes, beam current, parameter limit, etc.
\item  Read the monitor value applying moving average, limit check and 
  other specific post-processing.
\item  Derive the feedback amount applying conversion factors, 
  gain and limit check.
\item  Set the actuator value applying limit check and other 
  specific pre-processing.
\item  Flow control, graphics display, recording and interface to 
  other software. 
\end{Itemize}

Fig.~\ref{fig-efb} shows an example panel for one of the energy feedback loops.
Each parameter in the panel can be modified at any time. 

Energy feedback loops are composed of a monitor of a beam position at
a large dispersion and an actuator of rf phases at two adjacent klystron
stations, in order to maintain a small energy spread.  This type of
energy feedback is installed at 4 locations at 180-degree arc and the
end of the linac.  Some parameters are different, depending on the beam
modes\cite{lin-efb-ical99}.

\begin{figure}[bht]
\centering
\includegraphics*[width=82mm]{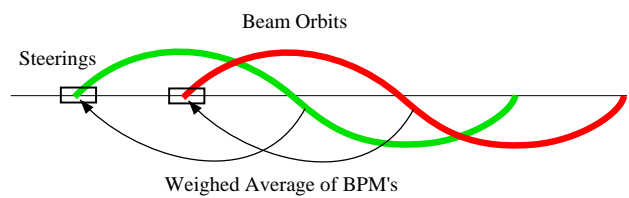}
\caption{Group of orbit feedback loops.  The weighed average of BPM's over 
a betatron wavelength as a response function is fed into a steering 
magnet strength.}
\label{fig-orb}
\end{figure}

Orbit feedback loops use beam positions as monitor values and steering
magnets as actuators.  A monitor value is actually a weighed average
of beam position monitors (BPM's) over a betatron wavelength according
to the response function for the corresponding steering magnet.
Normally, two subsequent regions, which are apart by 90-degree betatron
phase, are grouped as in Fig.~\ref{fig-orb}. Some feedback loops read
only one BPM, and are used to maintain the beam position at the end of the
linac.

This type of orbit feedback loop is installed at many sectors, and 
the current number of groups has reached 15.  

Since BPM's can be read at 1 Hz, most of the beam feedback loops are
operated at this speed\cite{lin-bpm-daq}.  The feedback gains are chosen
to be small, 0.2 to 0.5, in order to avoid oscillation.

The same feedback software has been applied to stabilize the accelerator
equipment.  Although these loops may be gradually moved to local
controllers, or even to the hardware, they are effective to suppress newly
found instabilities in the accelerator equipment.  Recently, it was
applied to suppress a long-term drift of the trigger timing of the
electron gun, and were found to be effective\cite{preinjector-epac2000}.

New feedback loops can be easily built simply by defining monitors,
actuators and some parameters. Standard software libraries provide 
an environment to tune those new loops, even during operation. 

In order to manage a large number of feedback loops, several software
panels were developed, such as a global feedback status display and
a feedback history viewer. 

\section{Conclusions}

Software was developed to stabilize the linac beam, and successfully
improved the beam reproducibility and reliability. Since it was
designed to be re-arranged easily, operators could solve problems
by modifying the software parameters, even when a beam operation mode had to
be modified.

The beam-mode switching panel has become very reliable, and has reproduced
beams sufficiently with switching more than 50 times a day. The switching
time, which is important for the integrated luminosity, was shorten to be
90 to 120 seconds. 

Feedback loops cured both the short-term instabilities and long-term
drifts of the beam energy, orbits and equipment parameters.  Depending
on the accelerator status, they suppressed the beam instability to one
half and drifts to one fifth without any operator manipulations. It was also 
useful to keep beams when beam studies were carried under unusual 
beam conditions and to find some anomalies in the accelerator. 

Those software systems were used in routine operation and
contributed to enhance the KEKB experiment efficiency.

\end{document}